\documentclass[prd,onecolumn,groupedaddress,noshowpacs,nofootinbib]{revtex4}
\usepackage{graphicx}
\usepackage{dcolumn}
\usepackage{amssymb}
\usepackage{mathrsfs}
\usepackage{amsmath}
\usepackage{epsfig}
\usepackage[dvips]{color}
\usepackage{hhline}

\begin{document}

\title{Extending two Higgs doublet models for two-loop neutrino mass generation and one-loop neutrinoless double beta decay}

\author{Zhen Liu}

\email{liu-zhen@sjtu.edu.cn}

\author{Pei-Hong Gu}

\email{peihong.gu@sjtu.edu.cn}

\affiliation{Department of Physics and Astronomy, Shanghai Jiao Tong University, 800 Dongchuan Road, Shanghai 200240, China}

\begin{abstract}

We extend some two Higgs doublet models, where the Yukawa couplings for the charged fermion mass generation only involve one Higgs doublet, by two singlet scalars respectively carrying a singly electric charge and a doubly electric charge. The doublet and singlet scalars together can mediate a two-loop diagram to generate a tiny Majorana mass matrix of the standard model neutrinos. Remarkably, the structure of the neutrino mass matrix is fully determined by the symmetric Yukawa couplings of the doubly charged scalar to the right-handed leptons. Meanwhile, a one-loop induced neutrinoless double beta decay can arrive at a testable level even if the electron neutrino has an extremely small Majorana mass. We also study other experimental constraints and implications including some rare processes and Higgs phenomenology.

\end{abstract}

\maketitle

\section{Introduction}

The massive and mixing neutrinos have been confirmed by the precise measurements on the atmospheric, solar, accelerator and reactor neutrino oscillations \cite{patrignani2016}. This fact implies the need for new physics beyond the $SU(3)_c^{}\times SU(2)_L^{}\times U(1)^{}_{Y}$ standard model (SM). On the other hand, the cosmological observations have indicated that the neutrino masses should be in a sub-eV range \cite{patrignani2016}. In order to understand the small neutrino masses, people have proposed various ideas, among which the tree-level seesaw \cite{minkowski1977,yanagida1979,grs1979,ms1980} mechanism is very popular \cite{minkowski1977,yanagida1979,grs1979,ms1980,mw1980,sv1980,cl1980,lsw1981,ms1981,flhj1989,ma1998}. However, the seesaw will not be easy to verify unless it is not at a naturally high scale. Alternatively, the neutrinos can acquire their tiny masses at loop order \cite{zee1980,zee1985,babu1988,bl2001,knt2003,cs2004,ma2006,lln2009,kss2011,acmn2014,amn2014,addh2015,hhop2015,ma2015,lg2015,amn2016,ammn2016,noo2016-3,no2016,no2016-2,gms2016,noo2016,ghl2016,noo2016-2,cgn2006,cgnw2006,cg2010,dabsw2012,gnr2013,cght2013,gnr2014,ght2014,gt2016}. These models for the radiative neutrino mass generation contain additionally charged scalars so that they may be tested at colliders.

In principle the neutrinos can have a Majorana nature \cite{majorana1937} since they do not carry any electric charges. One hence can expect a neutrinoless double beta decay ($0\nu\beta\beta$) \cite{furry1939} process mediated by the Majorana electron neutrinos. This $0\nu\beta\beta$ process is determined by one unknown parameter $m_{ee}^{}$, i.e. the $1\!-\!1$ element in the Majorana neutrino mass matrix, so that it can be seen in the running and planning experiments unless the $m_{ee}^{}$ parameter is big enough \cite{rodejohann2011,bgmr2013}. However, there are other possibilities for a $0\nu\beta\beta$ process \cite{ms1980,ms1981,tnnsv2010,pp2012,ddkv2012,app2013,br2013,hl2014,dgpss2015,bgm2015,glp2015,abm2015,bbgm2015}. For example, some left-right symmetric models for a linear seesaw of tree-level neutrino mass generation can of\mbox{}fer a nonconventional tree-level $0\nu\beta\beta$ process with an testable lifetime, which simply depends on the scale of the left-right symmetry breaking rather than the details of the Majorana neutrino mass matrix \cite{gu2015}. One may also consider other models which accommodate an observable $0\nu\beta\beta$ process at tree level and then give a negligible contribution to the neutrino masses at loop order \cite{sv1982}. These $0\nu\beta\beta$ processes, which are related to quite a few arbitrary parameters, thus can be free of the constraint from the neutrino mass matrix \cite{bm2002,gu2011,bhow2013}.

It should be interesting if an enhanced $0\nu\beta\beta$ process originates from a tiny $m_{ee}^{}$. Some people have realized this scenario \cite{cgn2006,cgnw2006,cg2010,dabsw2012,gnr2013,cght2013,gnr2014,ght2014,gt2016}. In a realistic model \cite{cgn2006}, after the SM Higgs doublet develops a vacuum expectation value (VEV) for spontaneously breaking the electroweak symmetry, a Higgs triplet without any Yukawa couplings can acquire an induced VEV up to a few GeV, meanwhile, its doubly charged component can mix with a doubly charged scalar singlet. Thanks to the gauge interactions, a two-loop induced Majorana neutrino mass matrix then can have a structure fully determined by the symmetric Yukawa couplings of the doubly charged scalar singlet to the right-handed leptons. As for the $0\nu\beta\beta$ process, it can appear at tree level through the same Yukawa interactions and the related gauge interactions.

In this paper we shall extend some two Higgs doublet models \cite{bflrss2012}, where the Yukawa couplings for the charged fermion mass generation only involve one Higgs scalar, to generate the required neutrino masses and the enhanced $0\nu\beta\beta$ process. Specifically we shall introduce two scalar singlets among which one carries a singly electric charge while the other one carries a doubly electric charge. The singly charged scalar without any Yukawa couplings has a cubic term with the two Higgs scalars. The doubly charged scalar has the Yukawa couplings with the right-handed leptons, besides its trilinear coupling with the singly charged scalar. After the electroweak symmetry breaking, we can obtain a dominant Majorana neutrino mass matrix at two-loop and a negligible one at three-loop level. The symmetric Yukawa couplings of the doubly charged scalar to the right-handed leptons can fully determine the structure of the neutrino mass matrix. The $0\nu\beta\beta$ processes can be induced at tree, one-loop and two-loop level. The amplitudes of these $0\nu\beta\beta$ processes are all proportional to the electron neutrino mass. The one-loop $0\nu\beta\beta$ process can arrive at an observable level even if the electron neutrino mass is extremely small. We will also study other experimental constraints and implications including some rare processes and Higgs phenomenology.

\section{The models}

In the fermion sector, the quarks and leptons are as same as the SM ones,
\begin{eqnarray}
&&\begin{array}{l}q^{}_L(3,2,+\frac{1}{6})\end{array}\!=\!\left[\begin{array}{l}u^{}_L\\
[1mm]
d^{}_L\end{array}\right],~\begin{array}{l}d^{}_R(3,1,-\frac{1}{3})\end{array},
~\begin{array}{l}u^{}_R(3,1,+\frac{2}{3})\end{array},~\begin{array}{l}l^{}_L(1,2,-\frac{1}{2})\end{array}\!=\!\left[\begin{array}{l}\nu^{}_L\\
[1mm]
e^{}_L\end{array}\right],~\begin{array}{l}e^{}_R(1,1,-1)\end{array}\,.
\end{eqnarray}
Here and thereafter the brackets following the fields describe the transformations under the $SU(2)^{}_c\times SU(2)^{}_L\times U(1)^{}_{Y}$ gauge groups. The scalar sector contains two charged singlets,
\begin{eqnarray}
\begin{array}{l}\delta^{\pm}_{}(1,1,\pm1)\end{array},
~\begin{array}{l}\xi^{\pm\pm}_{}(1,1,\pm2)\end{array}\,,
\end{eqnarray}
besides two Higgs doublets,
\begin{eqnarray}
\begin{array}{l}\phi^{}_1(1,2,+\frac{1}{2})\end{array}\!=\!\left[\begin{array}{l}\phi^{+}_{1}\\
[2mm]
\phi^{0}_{1}\end{array}\right],
~\begin{array}{l}\phi^{}_2(1,2,+\frac{1}{2})\end{array}\!=\!\left[\begin{array}{l}\phi^{+}_{2}\\
[2mm]
\phi^{0}_{2}\end{array}\right]\,.
\end{eqnarray}

We impose a softly broken discrete symmetry $\mathbb{S}$ so that the Yukawa couplings for generating the charged fermion masses will only involve one Higgs doublet. For example, we can take $\mathbb{S}=Z_2^{}$ under which one type or three types of the right-handed fermions and one Higgs scalar carry an odd parity while the other fermions and scalars carry an even parity. Alternatively, the $\mathbb{S}$ symmetry can be a global one. For instance,  we can take $\mathbb{S}=U(1)_X^{}$ under which only one Higgs scalar is non-trivial. We further assume a softly broken lepton number, under which the doubly charged scalar carries a lepton number of two units while the Higgs scalars and the singly charged scalar do not carry any lepton numbers.

We then summarize the allowed Yukawa interactions as below,
\begin{itemize}
\item \emph{Case-1}: The right-handed up-type quarks, down-type quarks and charged leptons couple to a same Higgs doublet,
\begin{eqnarray}
\mathcal{L}_Y^{} &=&-y'^{}_u\bar{q}_L^{}\tilde{\phi}^{}_1 u_R^{}-y'^{}_d\bar{q}_L^{}\phi^{}_1 d_R^{}
-y'^{}_e\bar{l}_L^{}\phi^{}_1 e_R^{}-\frac{1}{2}f\xi^{--}_{}\bar{e}_R^{}e^{c}_R+\textrm{H.c.}~~\textrm{with}~~f=f^T_{}\,;
\end{eqnarray}
\item \emph{Case-2}: The right-handed up-type quarks and  down-type quarks couple to a same Higgs doublet,
\begin{eqnarray}
\mathcal{L}_Y^{} &=&-y'^{}_u\bar{q}_L^{}\tilde{\phi}^{}_1 u_R^{}-y'^{}_d\bar{q}_L^{}\phi^{}_1 d_R^{}
-y'^{}_e\bar{l}_L^{}\phi^{}_2 e_R^{}-\frac{1}{2}f\xi^{--}_{}\bar{e}_R^{}e^{c}_R+\textrm{H.c.}~~\textrm{with}~~f=f^T_{}\,;
\end{eqnarray}
\item \emph{Case-3}: The right-handed up-type quarks and charged leptons couple to a same Higgs doublet,
\begin{eqnarray}
\mathcal{L}_Y^{} &=&-y'^{}_u\bar{q}_L^{}\tilde{\phi}^{}_1 u_R^{}-y'^{}_d\bar{q}_L^{}\phi^{}_2 d_R^{}
-y'^{}_e\bar{l}_L^{}\phi^{}_1 e_R^{}-\frac{1}{2}f\xi^{--}_{}\bar{e}_R^{}e^{c}_R+\textrm{H.c.}~~\textrm{with}~~f=f^T_{}\,;
\end{eqnarray}
\item \emph{Case-4}: The right-handed down-type quarks and charged leptons couple to a same Higgs doublet,
\begin{eqnarray}
\mathcal{L}_Y^{} &=&-y'^{}_u\bar{q}_L^{}\tilde{\phi}^{}_1 u_R^{}-y'^{}_d\bar{q}_L^{}\phi^{}_2 d_R^{}
-y'^{}_e\bar{l}_L^{}\phi^{}_2 e_R^{}-\frac{1}{2}f\xi^{--}_{}\bar{e}_R^{}e^{c}_R+\textrm{H.c.}~~\textrm{with}~~f=f^T_{}\,.
\end{eqnarray}
\end{itemize}

We also write down the full scalar potential with the following quadratic, cubic and quartic terms,
\begin{eqnarray}
\label{potential}
V&=&\mu_1^2|\phi_1^{}|^2_{}
+\mu_2^2 |\phi_2^{}|^2_{}+\mu_{12}^2(\phi^\dagger_1\phi^{}_2+\textrm{H.c.})
+\lambda_1^{}|\phi_1^{}|^4_{}+\lambda_2^{}|\phi_2^{}|^4_{}
+\lambda_{12}^{}|\phi_1^{}|^2_{}|\phi_2^{}|^2_{}
+\lambda'^{}_{12}\phi_1^\dagger\phi_2^{}\phi_2^\dagger\phi_1^{}\nonumber\\
[2mm]&&+[\lambda''^{}_{12}(\phi_1^\dagger\phi_2^{})^2_{}+\textrm{H.c.}]+(\mu_\delta^2
+\lambda_{1\delta}^{}|\phi_1^{}|^2_{}+\lambda_{2\delta}^{}|\phi_2^{}|^2_{})|\delta|^2_{}
+\lambda_\delta^{}|\delta|^4_{}+(\mu_\xi^2
+\lambda_{1\xi}^{}|\phi_1^{}|^2_{}+\lambda_{2\xi}^{}|\phi_2^{}|^2_{})|\xi|^2_{}
+\lambda_\xi^{}|\xi|^4_{}\nonumber\\
[2mm]&&+\lambda_{\delta\xi}^{}|\delta|^2_{}|\xi|^2_{}+\frac{1}{2}\omega(\xi^{++}_{}\delta^{-}_{}\delta^{-}_{}+\textrm{H.c.})+\sigma_{12}^{}(\delta^{-}_{}\phi^T_1 i\tau_2^{}\phi_2^{}+\textrm{H.c.})\,.\end{eqnarray}
Note only the $\mu_{12}^2$-term, the $\omega$-term and the $\sigma_{12}^{}$-term can softly break the additional $\mathbb{S}$ symmetry. So, if the $\mathbb{S}$ symmetry is a global one, the $\lambda''^{}_{12}$-term should be absent.

\section{Electroweak symmetry breaking and physical scalars}

The two Higgs scalars $\phi_{1,2}^{}$ can be always rotated by
\begin{eqnarray}
\chi=\left[\begin{array}{c}\chi^{+}_{}\\
[1mm]
\chi^{0}_{}\end{array}\right]=\phi_1^{}\cos\beta-\phi_2^{}\sin\beta\,,\quad
\varphi=\left[\begin{array}{c}\varphi^{+}_{}\\
[1mm]
\varphi^{0}_{}\end{array}\right]=\phi_1^{}\sin\beta+\phi_2^{}\cos\beta\,.
\end{eqnarray}
For a proper choice of the rotation angle $\beta$, only one of the newly defined Higgs scalars $\chi$ and $\varphi$ will develop a nonzero VEV to spontaneously break the electroweak symmetry. Without loss of generality, we can denote
\begin{eqnarray}
\tan\beta=\frac{\langle\phi_1^{}\rangle}{\langle\phi_2^{}\rangle}
=\frac{v_1^{}}{v_2^{}}
~~\textrm{with}~~\langle\phi_{1}^{}\rangle=\left[\begin{array}{c}0\\
[2mm]
\frac{1}{\sqrt{2}}v^{}_{1}\end{array}\right]\,,~~\langle\phi_{2}^{}\rangle=\left[\begin{array}{c}0\\
[2mm]
\frac{1}{\sqrt{2}}v^{}_{2}\end{array}\right]\,,
\end{eqnarray}
and then take
\begin{eqnarray}
\langle\chi\rangle=0\,,\quad\langle\varphi\rangle=\left[\begin{array}{c}0\\
[2mm]
\frac{1}{\sqrt{2}}v^{}_{}\end{array}\right]~~\textrm{with}~~v=\sqrt{v_1^2+v_2^2}\simeq 246\,\textrm{GeV}\,.\quad\quad
\end{eqnarray}
This means the Higgs scalar $\varphi$ will be responsible for the electroweak symmetry breaking.

In the base with $\langle\chi\rangle=0$, we can rewrite the scalar potential to be
\begin{eqnarray}
\label{potential2}
V&=&\mu_\varphi^2|\varphi|^2_{}+M_\chi^2 |\chi|^2_{}
+\kappa_1^{}|\varphi|^4_{}+\kappa_2^{}|\chi|^4_{}
+\kappa_{3}^{}|\varphi|^2_{}|\chi|^2_{}+\kappa^{}_{4}\varphi_{}^\dagger\chi\chi_{}^\dagger\varphi
+\kappa^{}_{5}[(\varphi_{}^\dagger\chi)^2_{}+\textrm{H.c.}]+(M_\delta^2
+\kappa_{\varphi\delta}^{}|\varphi|^2_{}\nonumber\\
[2mm]&&+\kappa_{\chi\delta}^{}|\chi|^2_{}+\kappa_{\varphi\chi\delta}^{}\varphi^\dagger_{}\chi+\textrm{H.c.})|\delta|^2_{}
+\kappa_\delta^{}|\delta|^4_{}+(M_\xi^2+\kappa_{\varphi\xi}^{}|\varphi|^2_{}
+\kappa_{\chi\xi}^{}|\chi|^2_{}+\kappa_{\varphi\chi\xi}^{}\varphi^\dagger_{}\chi+\textrm{H.c.})|\xi|^2_{}+\kappa_\xi^{}|\xi|^4_{}\nonumber\\
[2mm]&&
+\kappa_{\delta\xi}^{}|\delta|^2_{}|\xi|^2_{}
+\frac{1}{2}\omega(\xi^{++}_{}\delta^{-}_{}\delta^{-}_{}+\textrm{H.c.})+\sigma(\delta^{-}_{}\varphi^T_{} i\tau_2^{}\chi+\textrm{H.c.})\,.
\end{eqnarray}
Meanwhile, the Yukawa interactions can be expanded by
\begin{itemize}
\item \emph{Case-1}: 
\begin{eqnarray}
\mathcal{L}_Y^{} &=&-y^{}_u\bar{q}_L^{}\tilde{\varphi} u_R^{}-y^{}_d\bar{q}_L^{}\varphi d_R^{}
-y^{}_e\bar{l}_L^{}\varphi e_R^{}- (y^{}_u\cot\beta)\bar{q}_L^{}\tilde{\chi} u_R^{}-(y^{}_d\cot\beta) \bar{q}_L^{}\chi d_R^{}-(y^{}_e\cot\beta) \bar{l}_L^{}\chi e_R^{}\nonumber\\
[2mm]&&-\frac{1}{2}f\xi^{--}_{}\bar{e}_R^{}e^{c}_R+\textrm{H.c.}\,;
\end{eqnarray}
\item \emph{Case-2}:
\begin{eqnarray}
\mathcal{L}_Y^{} &=&-y^{}_u\bar{q}_L^{}\tilde{\varphi} u_R^{}-y^{}_d\bar{q}_L^{}\varphi d_R^{}
-y^{}_e\bar{l}_L^{}\varphi e_R^{}- (y^{}_u\cot\beta)\bar{q}_L^{}\tilde{\chi} u_R^{}-(y^{}_d\cot\beta) \bar{q}_L^{}\chi d_R^{}+(y^{}_e\tan\beta) \bar{l}_L^{}\chi e_R^{}\nonumber\\
[2mm]&&-\frac{1}{2}f\xi^{--}_{}\bar{e}_R^{}e^{c}_R+\textrm{H.c.}\,;
\end{eqnarray}
\item \emph{Case-3}:
\begin{eqnarray}
\mathcal{L}_Y^{} &=&-y^{}_u\bar{q}_L^{}\tilde{\varphi} u_R^{}-y^{}_d\bar{q}_L^{}\varphi d_R^{}
-y^{}_e\bar{l}_L^{}\varphi e_R^{}- (y^{}_u\cot\beta)\bar{q}_L^{}\tilde{\chi} u_R^{}+(y^{}_d\tan\beta) \bar{q}_L^{}\chi d_R^{}
-(y^{}_e\cot\beta) \bar{l}_L^{}\chi e_R^{}\nonumber\\
[2mm]&&-\frac{1}{2}f\xi^{--}_{}\bar{e}_R^{}e^{c}_R+\textrm{H.c.}\,;
\end{eqnarray}
\item \emph{Case-4}:
\begin{eqnarray}
\mathcal{L}_Y^{} &=&-y^{}_u\bar{q}_L^{}\tilde{\varphi} u_R^{}-y^{}_d\bar{q}_L^{}\varphi d_R^{}
-y^{}_e\bar{l}_L^{}\varphi e_R^{}-(y^{}_u\cot\beta)\bar{q}_L^{}\tilde{\chi} u_R^{}+(y^{}_d\tan\beta) \bar{q}_L^{}\chi d_R^{}
+(y^{}_e\tan\beta) \bar{l}_L^{}\chi e_R^{}\nonumber\\
[2mm]&&-\frac{1}{2}f\xi^{--}_{}\bar{e}_R^{}e^{c}_R+\textrm{H.c.}\,.
\end{eqnarray}
\end{itemize}

After the Higgs scalar $\varphi$ develops its VEV for the electroweak symmetry breaking, we can take
\begin{eqnarray}
\varphi=\left[\begin{array}{c}0\\
[2mm]
\frac{1}{\sqrt{2}}(v+h)\end{array}\right],~~\chi=\left[\begin{array}{c}\chi^{+}_{}\\
[2mm]
\frac{1}{\sqrt{2}}(\chi^0_R+i\chi^0_I)\end{array}\right]\,.
\end{eqnarray}
The scalar potential (\ref{potential2}) then should give the mass terms as follows,
\begin{eqnarray}
\label{potential3}
V\supset\frac{1}{2}m_h^2 h^2_{}+\frac{1}{2}m_{\chi^{0}_R}^2(\chi^0_R)^2_{}+\frac{1}{2}m_{\chi^{0}_I}^2(\chi^0_I)^2_{}+m_{\chi^\pm_{}}^2 \chi^{+}_{}\chi^{-}_{}+m_{\delta^\pm_{}}^2 \delta^{+}_{}\delta^{-}_{}
+m_{\chi^\pm_{}\delta^\pm_{}}^2 (\chi^{+}_{}\delta^{-}_{}+\textrm{H.c.})+m_{\xi^{\pm\pm}_{}}^2 \xi^{++}_{}\xi^{--}_{}\,,
\end{eqnarray}
with
\begin{eqnarray}
m_h^2&=&2\kappa_1^{}v^2_{}=125\,\textrm{GeV}\,,\nonumber\\
[2mm]
m_{\chi^{0}_{R}}^2&=&M_\chi^2+\left(\frac{1}{2}\kappa_3^{}+\frac{1}{2}\kappa_4^{}+\kappa_5^{}\right)v^2_{}>0\,,\nonumber\\
[2mm]
m_{\chi^{0}_{I}}^2&=&M_\chi^2+\left(\frac{1}{2}\kappa_3^{}+\frac{1}{2}\kappa_4^{}-\kappa_5^{}\right)v^2_{}>0\,,\nonumber\\
[2mm]
m_{\chi^\pm_{}}^2&=&M_\chi^2+\frac{1}{2}\kappa_3^{}v^2_{}\,,\nonumber\\
[2mm]
m_{\delta^\pm_{}}^2&=&M_\delta^2+\frac{1}{2}\kappa_{\varphi\delta}^{}v^2_{}\,,\nonumber\\
[2mm]
m_{\chi^\pm_{}\delta^\pm_{}}^2&=&-\frac{1}{\sqrt{2}}v\sigma \,,\nonumber\\
[2mm]
m_{\xi^{\pm\pm}_{}}^2&=&M_\xi^2+\frac{1}{2}\kappa_{\varphi\xi}^{}v^2_{}>0\,.
\end{eqnarray}
Now the singly charged scalars $\delta^\pm_{}$ and $\chi^\pm_{}$ mix with each other. Their mass eigenstates should be
\begin{eqnarray}
\hat{\chi}^\pm_{}&=&\chi^\pm_{}\cos\theta-\delta^\pm_{}\sin\theta ~~\textrm{with}~~
m_{\hat{\chi}^\pm_{}}^2=\frac{m_{\chi^\pm_{}}^2+m_{\delta^\pm_{}}^2+\sqrt{(m_{\chi^\pm_{}}^2-m_{\delta^\pm_{}}^2)^2_{}+2v^2_{}\sigma^2_{}}}{2}>0\,,\nonumber\\
[2mm]
\hat{\delta}^\pm_{}&=&\chi^\pm_{}\sin\theta+\delta^\pm_{}\cos\theta ~~\textrm{with}~~
m_{\hat{\delta}^\pm_{}}^2=\frac{m_{\chi^\pm_{}}^2+m_{\delta^\pm_{}}^2-\sqrt{(m_{\chi^\pm_{}}^2-m_{\delta^\pm_{}}^2)^2_{}+2v^2_{}\sigma^2_{}}}{2}>0\,,
\end{eqnarray}
where the rotation angle $\theta$ is determined by
\begin{eqnarray}
\label{anglerotation}
\tan2\theta &=& \frac{\sqrt{2}v\sigma}{m_{\chi^{\pm}_{}}^2-m_{\delta^{\pm}_{}}^2}\Rightarrow 
\sin^2_{} 2\theta =\frac{2v^2_{}\sigma^2_{}}{(m_{\hat{\chi}^{\pm}_{}}^2-m_{\hat{\delta}^{\pm}_{}}^2)^2_{}}
=\frac{2\left(\frac{v^2_{}}{m_{\xi^{\pm\pm}_{}}^2}\right)\left(\frac{\sigma^2_{}}{m_{\xi^{\pm\pm}_{}}^2}\right)}{\left(\frac{m_{\hat{\chi}^{\pm}_{}}^2}{m_{\xi^{\pm\pm}_{}}^2}-\frac{m_{\hat{\delta}^{\pm}_{}}^2}{m_{\xi^{\pm\pm}_{}}^2}\right)^2_{}}\leq 1~~\textrm{for}\nonumber\\
[2mm]\Delta m_{\pm}^2&\equiv& m_{\hat{\chi}^{\pm}_{}}^2-m_{\hat{\delta}^{\pm}_{}}^2 =\sqrt{(m_{\chi^\pm_{}}^2-m_{\delta^\pm_{}}^2)^2_{}+2v^2_{}\sigma^2_{}}\geq  \sqrt{2}v |\sigma|\,.
\end{eqnarray}

The charged fermions can get their masses through the Yukawa couplings involving the Higgs scalar $\varphi$, i.e.  \cite{patrignani2016}
\begin{eqnarray}
\hat{m}_u^{}&=&\frac{v}{\sqrt{2}}\hat{y}_u^{}=\textrm{diag}\{m_u^{},~m_c^{},~m_t^{}\}\simeq\textrm{diag}\{2.2\,\textrm{MeV},~1.27\,\textrm{GeV},~173\,\textrm{GeV}\}\,,\nonumber\\
[2mm]
\hat{m}_d^{}&=&\frac{v}{\sqrt{2}}\hat{y}_d^{}=\textrm{diag}\{m_d^{},~m_s^{},~m_b^{}\}\simeq\textrm{diag}\{4.7\,\textrm{MeV},~96\,\textrm{MeV}\,,~4.18\,\textrm{GeV}\}\,,\nonumber\\
[2mm]
\hat{m}_e^{}&=&\frac{v}{\sqrt{2}}\hat{y}_e^{}=\textrm{diag}\{m_e^{},~m_\mu^{},~m_\tau^{}\}
\simeq\textrm{diag}\{0.511\,\textrm{MeV},~107\,\textrm{MeV},~1.78\,\textrm{GeV}\}\,.
\end{eqnarray}
Note the perturbation requirement $|\hat{y}'^{}_{t}|<\sqrt{4\pi}$, $|\hat{y}'^{}_{b}|<\sqrt{4\pi}$ and $|\hat{y}'^{}_{\tau}|<\sqrt{4\pi}$ will constrain,
\begin{itemize}
\item \emph{Case-1}:
\begin{eqnarray}
&&\cot\beta \lesssim 3.3~~\textrm{for}~~\cot\beta<\sqrt{\frac{2\pi v^2_{}}{m_t^2}-1}\,;
\end{eqnarray}
\item \emph{Case-2}:
\begin{eqnarray}
&&0.3\lesssim\tan\beta\lesssim 346~~\textrm{for}~~\frac{1}{\sqrt{\frac{2\pi v^2_{}}{m_t^2}-1}}<\tan\beta < \sqrt{\frac{2\pi v^2_{}}{m_\tau^2}-1}\,;
\end{eqnarray}
\item \emph{Case-3}:
\begin{eqnarray}
&&6.8\times 10^{-3}_{}\lesssim\cot\beta\lesssim 3.3~~\textrm{for}~~\frac{1}{ \sqrt{\frac{2\pi v^2_{}}{m_b^2}-1}}<\cot\beta <\sqrt{\frac{2\pi v^2_{}}{m_t^2}-1}\,;
\end{eqnarray}
\item \emph{Case-4}:
\begin{eqnarray}
&&0.3\lesssim\tan\beta\lesssim147~~\textrm{for}~~\frac{1}{\sqrt{\frac{2\pi v^2_{}}{m_t^2}-1}}<\tan\beta < \sqrt{\frac{2\pi v^2_{}}{m_b^2}-1}\,.
\end{eqnarray}
\end{itemize}
Moreover, the cubic couplings $\omega$ and $\sigma$ in the potential (\ref{potential2}) should also match the perturbation requirement. Roughly speaking, we have
\begin{eqnarray}
\omega\lesssim \textrm{max}\{M_\xi^{},~M_\delta^{}\}\,,~~\sigma\lesssim \textrm{max}\{M_\delta^{},~M_\chi^{}\}\,.
\end{eqnarray}

\section{Neutrino masses}

\begin{figure*}
\vspace{6.5cm} \epsfig{file=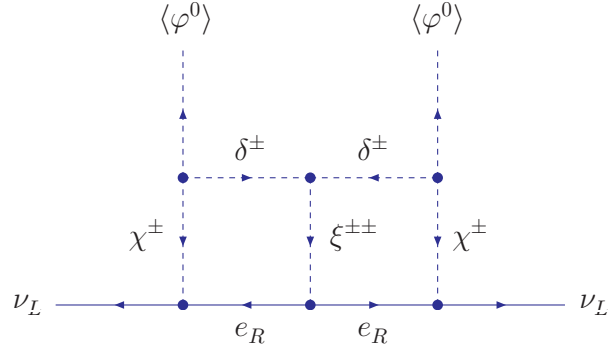, bbllx=6cm,
bblly=6.0cm, bburx=16cm, bbury=16cm, width=8cm, height=8cm,
angle=0, clip=0} \vspace{-9.25cm} \caption{\label{2loop}
Two-loop diagram for Majorana neutrino masses. }
\end{figure*}

\begin{figure*}
\vspace{7.5cm} \epsfig{file=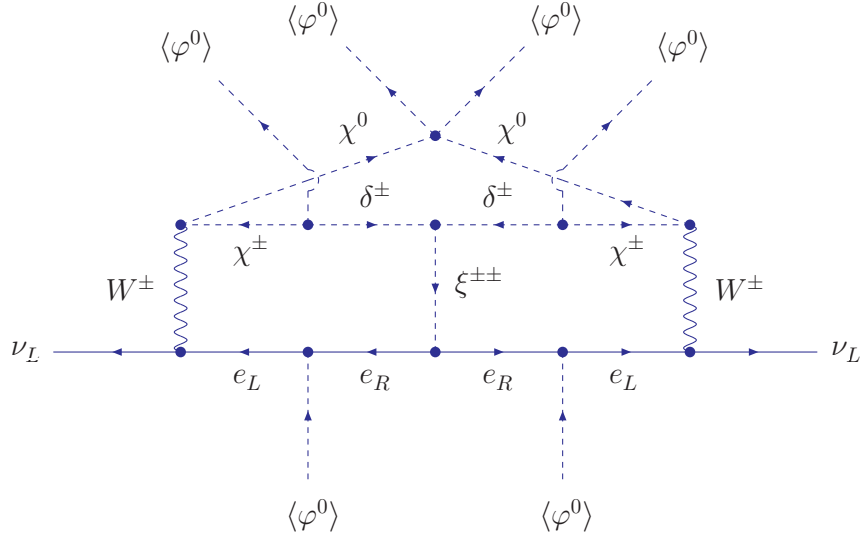, bbllx=6cm,
bblly=6.0cm, bburx=16cm, bbury=16cm, width=8cm, height=8cm,
angle=0, clip=0} \vspace{-7.5cm} \caption{\label{3loop}
Three-loop diagram for Majorana neutrino masses. }
\end{figure*}

As shown in Fig. \ref{2loop}, the left-handed neutrinos can obtain a Majorana mass term at two-loop level,
\begin{eqnarray}
\mathcal{L}&\supset&-\frac{1}{2}m_\nu^{}\bar{\nu}_L^{}\nu_L^c+\textrm{H.c.}\,.
\end{eqnarray}
It is easy to see that the two-loop neutrino mass matrix should have the structure as follows,
\begin{eqnarray}
\label{numass21}
m_\nu^{\textrm{2-loop}}\propto  \hat{y}_e^{}f\hat{y}_e^{}\propto \hat{m}_e^{}f\hat{m}_e^{}\,.
\end{eqnarray}
Although the three-loop diagram given in Fig. \ref{3loop} also generates the neutrino masses,
\begin{eqnarray}
\label{numass3}
m_\nu^{\textrm{3-loop}}\propto \hat{m}_e^{}f\hat{m}_e^{}\,,
\end{eqnarray}
its contribution should be much smaller than the two-loop contribution (\ref{numass21}). So, the neutrino mass matrix can be well described by
\begin{eqnarray}
\label{numass}
m_{\nu}^{}\simeq m_\nu^{\textrm{2-loop}}\gg m_\nu^{\textrm{3-loop}}\,.
\end{eqnarray}

\begin{figure*}[htp]
 \vspace{-1cm} \centering
  \includegraphics[width=15cm]{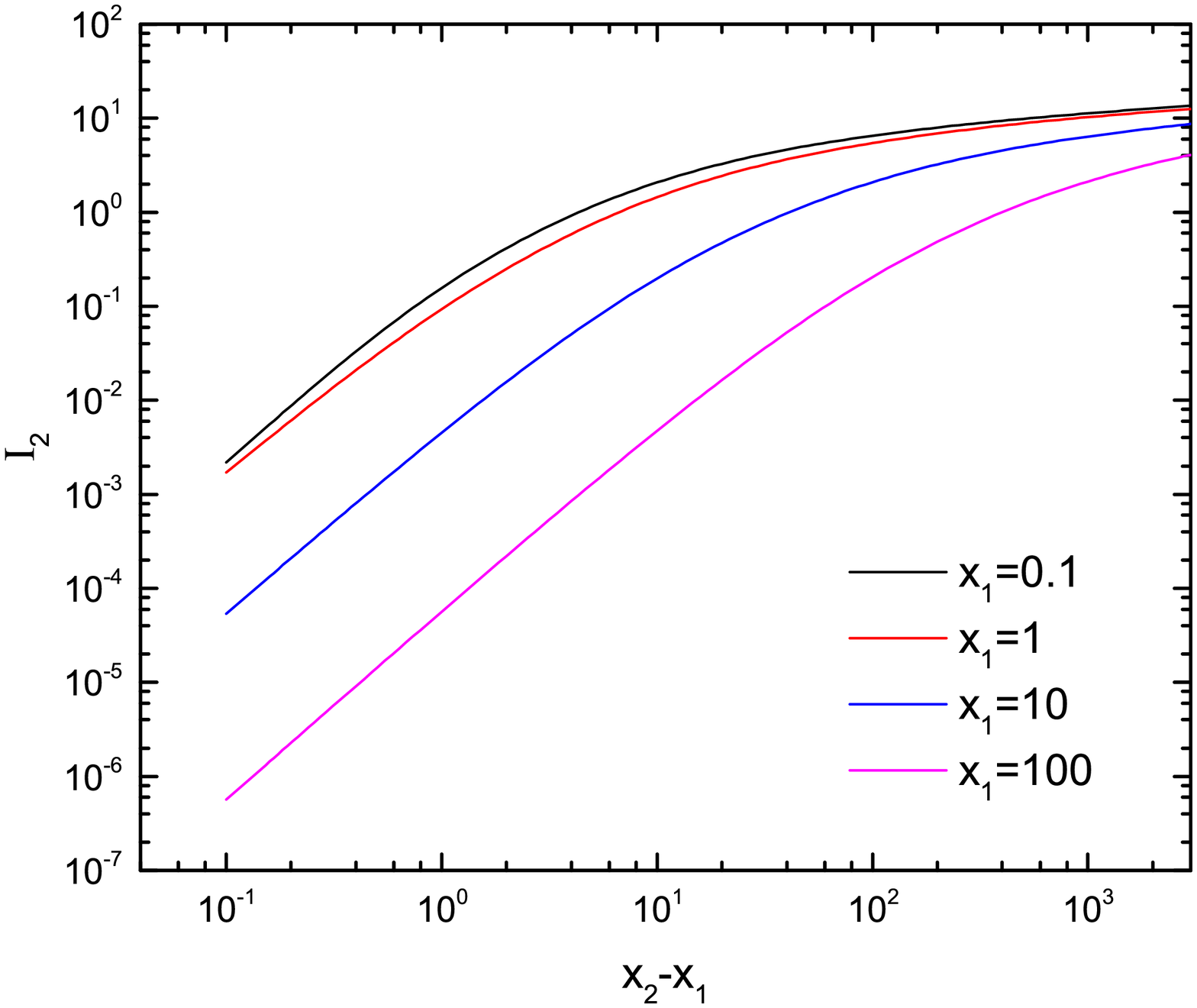}
  \caption{The two-loop integral $I_2^{}(x_1^{2},x_2^{2})$ as a function of the dif\mbox{}ference $x_2^{}-x_1^{}$ for a given $x_1^{}$. The lines from top to bottom correspond to $x_1^{}=0.1$, $x_1^{}=1$, $x_1^{}=10$ and $x_1^{}=100$, respectively. }
  \label{integral2}
\end{figure*}

We calculate
\begin{eqnarray}
\label{numass23}
m_\nu^{}=\frac{c I_2^{}\sin^2_{}2\theta}{2(16\pi^2_{})^2_{}} \hat{y}_e^{}f\hat{y}_e^{}\omega =\frac{ c I_2^{}\sin^2_{}2\theta}{2^{8}_{}\pi^4_{}}\frac{ \hat{m}_e^{}f \hat{m}_e^{}}{v^2_{}} \omega
\quad \textrm{with}\quad c=\left\{\begin{array}{ll}\cot^2_{}\beta& \textrm{in}~\emph{Case-1,3}\,,\\
[2mm]
\tan^2_{}\beta& \textrm{in}~\emph{Case-2,4}\,.
\end{array}\right.
\end{eqnarray}
Here $I_2^{}$ is given by the two-loop integral,
\begin{eqnarray}
I_2^{}\left(\frac{m_{\hat{\delta}^\pm_{}}^2}{m_{\xi^{\pm\pm}_{}}^2},\frac{m_{\hat{\chi}^\pm_{}}^2}{m_{\xi^{\pm\pm}_{}}^2}\right)\!&=&\!F_2^{}
\left(\frac{m_{\hat{\delta}^\pm_{}}^2}{m_{\xi^{\pm\pm}_{}}^2},\frac{m_{\hat{\delta}^\pm_{}}^2}{m_{\xi^{\pm\pm}_{}}^2}\right)+F_2^{}
\left(\frac{m_{\hat{\chi}^\pm_{}}^2}{m_{\xi^{\pm\pm}_{}}^2},\frac{m_{\hat{\chi}^\pm_{}}^2}{m_{\xi^{\pm\pm}_{}}^2}\right)
-2F_2^{}
\left(\frac{m_{\hat{\chi}^\pm_{}}^2}{m_{\xi^{\pm\pm}_{}}^2},\frac{m_{\hat{\delta}^\pm_{}}^2}{m_{\xi^{\pm\pm}_{}}^2}\right)\,,
\end{eqnarray}
where the function $F_2^{}(x_1^{2},x_2^{2})$ is defined by
\begin{eqnarray}
F_2^{}(x_1^{2},x_2^{2})=(16\pi^2_{})^2_{}\int\frac{d^4_{}\tilde{q}_1^{}}{(2\pi)^4_{}}\frac{d^4_{}\tilde{q}_2^{}}{(2\pi)^4_{}}\frac{/\!\!\!\tilde{q}_1^{}/\!\!\!\tilde{q}_2^{}}{\tilde{q}_1^2(\tilde{q}_1^2-x_1^{2})\tilde{q}_2^2(\tilde{q}_2^2-x_2^{2})
[(\tilde{q}_1^{}+\tilde{q}_2^{})^2_{}-1]}\,,
\end{eqnarray}
with $\tilde{q}_{1,2}^{}=q_{1,2}^{}/m_{\xi^{\pm\pm}_{}}^{}$ being the reduced momentum.

In Fig. \ref{integral2}, we show the numerical results of the two-loop integral $I_2^{}(x_1^2,x_2^2)$ as a function of the dif\mbox{}ference $ x_2^{}-x_1^{}$ for a given $x_1^{}$. The lines from top to bottom correspond to $x_1^{}=0.1$, $x_1^{}=1$, $x_1^{}=10$ and $x_1^{}=100$, respectively. For a proper parameter choice, the two-loop integral $I_2^{}$ can be of the order of $\mathcal{O}(1)$. Note that $I_2^{}(x_1^2,x_2^2)\equiv 0$ for $x_1^{}=x_2^{}$. We should also keep in mind that a bigger $x_2^{}-x_1^{}$ leads to a bigger $I_2^{}$ but a smaller $\sin^2_{}2\theta$, see Eq. (\ref{anglerotation}). The product $I_2^{}\sin^2_{}2\theta$ thus cannot be very large. We hence would fail in enhancing the neutrino masses by choosing a bigger cubic coupling $\omega$, which is not allowed to be much bigger than the charged scalar masses. In other words, the charged scalars cannot be far above the electroweak scale.

Now the neutrino mass matrix has a structure fully determined by the symmetric Yukawa couplings $f_{\alpha\beta}^{}=f_{\beta\alpha}^{}$ $(\alpha,\beta = e,\mu,\tau)$, i.e.
\begin{eqnarray}
\!\!\!\!m_{\alpha\beta}^{}&\equiv& (m_\nu^{})_{\alpha\beta}^{} =\frac{ c I_2^{}\sin^2_{}2\theta}{2^{8}_{}\pi^4_{}} \frac{\omega m_\alpha^{} m_\beta^{} }{v^2_{}}f_{\alpha\beta}^{} \,.
\end{eqnarray}
Actually we read
\begin{eqnarray}
m_{ee}^{}&=&6\times 10^{-4}\,\textrm{eV}\left(\frac{f_{ee}^{}}{\sqrt{4\pi}}\right) \left(\frac{c}{1}\right)\left(\frac{I_2^{}}{1}\right)\left(\frac{\sin^2_{} 2\theta}{1}\right)\left(\frac{\omega}{1\,\textrm{TeV}}\right) \,,\nonumber\\
[2mm]
m_{\mu\mu}^{}&=&0.1\,\textrm{eV}\left(\frac{f_{\mu\mu}^{}}{0.013}\right)\left(\frac{c}{1}\right)\left(\frac{I_2^{}}{1}\right)\left(\frac{\sin^2_{} 2\theta}{1}\right)\left(\frac{\omega}{1\,\textrm{TeV}}\right) \,,\nonumber\\
[2mm]
m_{\tau\tau}^{}&=&0.1\,\textrm{eV}\left(\frac{f_{\tau\tau}^{}}{5\times 10^{-5}_{}}\right)\left(\frac{c}{1}\right)\left(\frac{I_2^{}}{1}\right)\left(\frac{\sin^2_{} 2\theta}{1}\right)\left(\frac{\omega}{1\,\textrm{TeV}}\right) \,,\nonumber\\
[2mm]
m_{e\mu}^{}&=&m_{\mu e}^{}=0.1\,\textrm{eV}\left(\frac{f_{e\mu}^{}}{2.9}\right) \left(\frac{c}{1}\right)\left(\frac{I_2^{}}{1}\right)\left(\frac{\sin^2_{} 2\theta}{1}\right)\left(\frac{\omega}{1\,\textrm{TeV}}\right) \,,\nonumber\\
[2mm]
m_{e\tau}^{}&=&m_{\tau e}^{}=0.1\,\textrm{eV}\left(\frac{f_{e\tau}^{}}{0.17}\right)\left(\frac{c}{1}\right)\left(\frac{I_2^{}}{1}\right)\left(\frac{\sin^2_{} 2\theta}{1}\right)\left(\frac{\omega}{1\,\textrm{TeV}}\right) \,,\nonumber\\
[2mm]
m_{\mu\tau}^{}&=&m_{\tau \mu}^{}=0.1\,\textrm{eV}\left(\frac{f_{\mu\tau}^{}}{8\times 10^{-4}_{}}\right)\left(\frac{c}{1}\right)\left(\frac{I_2^{}}{1}\right)\left(\frac{\sin^2_{} 2\theta}{1}\right)\left(\frac{\omega}{1\,\textrm{TeV}}\right) \,.
\end{eqnarray}

In turn, we can parametrize the Yukawa couplings $f$ by the neutrino mass matrix,
\begin{eqnarray}
f&=&\frac{2^{8}_{}\pi^4_{}v^2_{}}{ c I_2^{} \sin^2_{}2\theta}\frac{1}{\hat{m}_e^{}}m_\nu^{} \frac{1}{\hat{m}_e^{}}\frac{1}{\omega}\,.
\end{eqnarray}

\section{Neutrinoless double beta decay}

It is well known that the electron neutrino with a Majorana mass can mediate a $0\nu\beta\beta$ process.
This most popular $0\nu\beta\beta$ picture can be described by the ef\mbox{}fective operator as below,
\begin{widetext}
\begin{eqnarray}
\label{eff2loop}
\mathcal{O}^{\textrm{2-loop}}_{}&=& \frac{g^4_{}}{2m_W^4}\bar{u}_L^{}\gamma^\mu_{}d_L^{}
 \bar{u}_L^{}\gamma^\nu_{}d_L^{}\bar{e}_L^{}\gamma_\mu^{}\gamma_\nu^{} e_L^c\frac{m_{ee}^{}}{q^2_{}}
 =\frac{1}{\Lambda_2^5}\bar{u}_L^{}\gamma^\mu_{}d_L^{}
 \bar{u}_L^{}\gamma^\nu_{}d_L^{}\bar{e}_L^{}\gamma_\mu^{}\gamma_\nu^{} e_L^c~~\textrm{with}~~
 \frac{1}{\Lambda_2^5}=16 \,G_F^2 \frac{m_{ee}^{}}{q^2_{}}\,,
\end{eqnarray}
\end{widetext}
where $q=100-200\,\textrm{MeV}$ is the transfer momentum. In the present model, the above $0\nu\beta\beta$ process actually is a two-loop ef\mbox{}fect since the neutrino masses are induced at two-loop level.

\begin{figure*}
\vspace{10cm} \epsfig{file=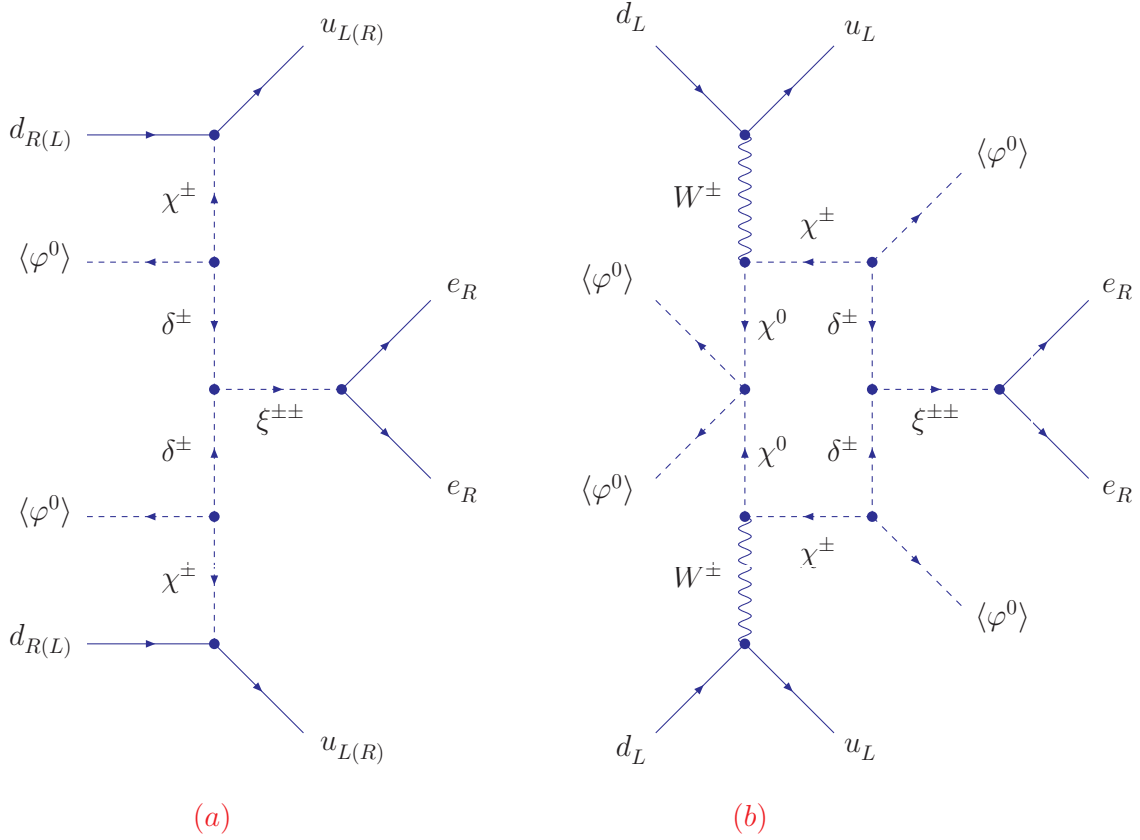, bbllx=5cm,
bblly=6.0cm, bburx=15cm, bbury=16cm, width=8cm, height=8cm,
angle=0, clip=0} \vspace{-6cm} \caption{\label{nuless}
Tree and one-loop diagrams for neutrinoless double beta decay. }
\end{figure*}

As shown in Figs. \ref{nuless}a, our model can also generate the other $0\nu\beta\beta$ processes at tree level. The ef\mbox{}fective operators should be
\begin{widetext}
\begin{eqnarray}
\label{efftree}
\mathcal{O}^{\textrm{tree}}_{}&=&\frac{1}{\Lambda_{01}^{5}}\bar{u}_L^{}d_R^{}\bar{u}_L^{}d_R^{}\bar{e}_R^c e_R^{}+\frac{1}{\Lambda_{02}^{5}}\bar{u}_R^{}d_L^{}\bar{u}_R^{}d_L^{}\bar{e}_R^c e_R^{}+\frac{1}{\Lambda_{03}^{5}}\bar{u}_L^{}d_R^{}\bar{u}_R^{}d_L^{}\bar{e}_R^c e_R^{}~~\textrm{with}\nonumber\\
[2mm]
\frac{1}{\Lambda_{01}^{5}}&=&\frac{1}{4}\hat{y}_d^2  f_{ee}^{} c^{}_{1}\sin^2_{}2\theta\frac{\omega}{m^{2}_{\xi^{\pm\pm}_{}}}\left(\frac{1}{m_{\hat{\chi}^{\pm}_{}}^2}-\frac{1}{m_{\hat{\delta}^{\pm}_{}}^2}\right)^2_{}
=\frac{2^7_{}\pi^4_{}}{I_2^{}}\frac{c_1^{}}{c}\frac{m_d^2}{m_e^2}\frac{m_{ee}^{}\Delta m_{\pm}^2}{m^{2}_{\xi^{\pm\pm}_{}} m_{\hat{\chi}^{\pm}_{}}^2 m_{\hat{\delta}^{\pm}_{}}^2}\nonumber\\
[2mm]
&=&2\times \left(\frac{c_1^{}}{c}\right)\left(\frac{m_d^{}}{3\,\textrm{MeV}}\right)^2_{}\left(\frac{q}{100\,\textrm{MeV}}\right)^2_{}\left(\frac{\Delta m_{\pm}^2}{1\,\textrm{TeV}^2_{}}\right)\left(\frac{1\,\textrm{TeV}}{m_{\xi^{\pm\pm}_{}}^{}}\right)^2_{}
\left(\frac{1\,\textrm{TeV}}{m_{\hat{\chi}^{\pm}_{}}^{}}\right)^2_{}\left(\frac{1\,\textrm{TeV}}{m_{\hat{\delta}^{\pm}_{}}^{}}\right)^2_{}\frac{1}{\Lambda_2^5}
\,,\nonumber\\
[2mm]
\frac{1}{\Lambda_{02}^{5}}&=&\frac{1}{4}\hat{y}_u^2  f_{ee}^{} c^{}_{2}\sin^2_{}2\theta\frac{\omega}{m^{2}_{\xi^{\pm\pm}_{}}}\left(\frac{1}{m_{\hat{\chi}^{\pm}_{}}^2}-\frac{1}{m_{\hat{\delta}^{\pm}_{}}^2}\right)^2_{}
=\frac{2^7_{}\pi^4_{}}{I_2^{}}\frac{c_2^{}}{c}\frac{m_u^2}{m_e^2}\frac{m_{ee}^{}\Delta m_{\pm}^2}{m^{2}_{\xi^{\pm\pm}_{}} m_{\hat{\chi}^{\pm}_{}}^2 m_{\hat{\delta}^{\pm}_{}}^2}\nonumber\\
[2mm]
&=&0.5\times \left(\frac{c_2^{}}{c}\right)\left(\frac{m_u^{}}{1.5\,\textrm{MeV}}\right)^2_{}\left(\frac{q}{100\,\textrm{MeV}}\right)^2_{}\left(\frac{\Delta m_{\pm}^2}{1\,\textrm{TeV}^2_{}}\right)\left(\frac{1\,\textrm{TeV}}{m_{\xi^{\pm\pm}_{}}^{}}\right)^2_{}
\left(\frac{1\,\textrm{TeV}}{m_{\hat{\chi}^{\pm}_{}}^{}}\right)^2_{}\left(\frac{1\,\textrm{TeV}}{m_{\hat{\delta}^{\pm}_{}}^{}}\right)^2_{}
\frac{1}{\Lambda_2^5}\,,\nonumber\\
[2mm]
\frac{1}{\Lambda_{03}^{5}}&=&\frac{1}{4}\hat{y}_d^{}\hat{y}_u^{}  f_{ee}^{} c^{}_{3}\sin^2_{}2\theta\frac{\omega}{m^{2}_{\xi^{\pm\pm}_{}}}\left(\frac{1}{m_{\hat{\chi}^{\pm}_{}}^2}-\frac{1}{m_{\hat{\delta}^{\pm}_{}}^2}\right)^2_{}
=\frac{2^7_{}\pi^4_{}}{I_2^{}}\frac{c_3^{}}{c}\frac{m_d^{}m_u^{}}{m_e^2}\frac{m_{ee}^{}\Delta m_{\pm}^2}{m^{2}_{\xi^{\pm\pm}_{}} m_{\hat{\chi}^{\pm}_{}}^2 m_{\hat{\delta}^{\pm}_{}}^2}\nonumber\\
[2mm]
&=&1\times \left(\frac{c_3^{}}{c}\right) \left(\frac{m_d^{}}{3\,\textrm{MeV}}\right)\left(\frac{m_u^{}}{1.5\,\textrm{MeV}}\right)\left(\frac{q}{100\,\textrm{MeV}}\right)^2_{}\left(\frac{\Delta m_{\pm}^2}{1\,\textrm{TeV}^2_{}}\right)\left(\frac{1\,\textrm{TeV}}{m_{\xi^{\pm\pm}_{}}^{}}\right)^2_{}
\left(\frac{1\,\textrm{TeV}}{m_{\hat{\chi}^{\pm}_{}}^{}}\right)^2_{}\left(\frac{1\,\textrm{TeV}}{m_{\hat{\delta}^{\pm}_{}}^{}}\right)^2_{}
\frac{1}{\Lambda_2^5}\,,
\end{eqnarray}
\end{widetext}
where the coef\mbox{}ficients $c_{1,2,3}^{}$ are defined by
\begin{eqnarray}
(c_1^{},c_2^{},c_3^{})=\left\{\begin{array}{ll}(c,c,c)& \textrm{in}~\emph{Case-1}\,,\\
[2mm]
(\frac{1}{c},\frac{1}{c},\frac{1}{c})& \textrm{in}~\emph{Case-2}\,,\\[2mm]
(\frac{1}{c},c,1)& \textrm{in}~\emph{Case-3}\,,\\[2mm]
(c,\frac{1}{c},1)& \textrm{in}~\emph{Case-4}\,.\\[2mm]
\end{array}\right.
\end{eqnarray}

\begin{figure*}[htp]
 \vspace{0cm} \centering
  \includegraphics[width=15cm]{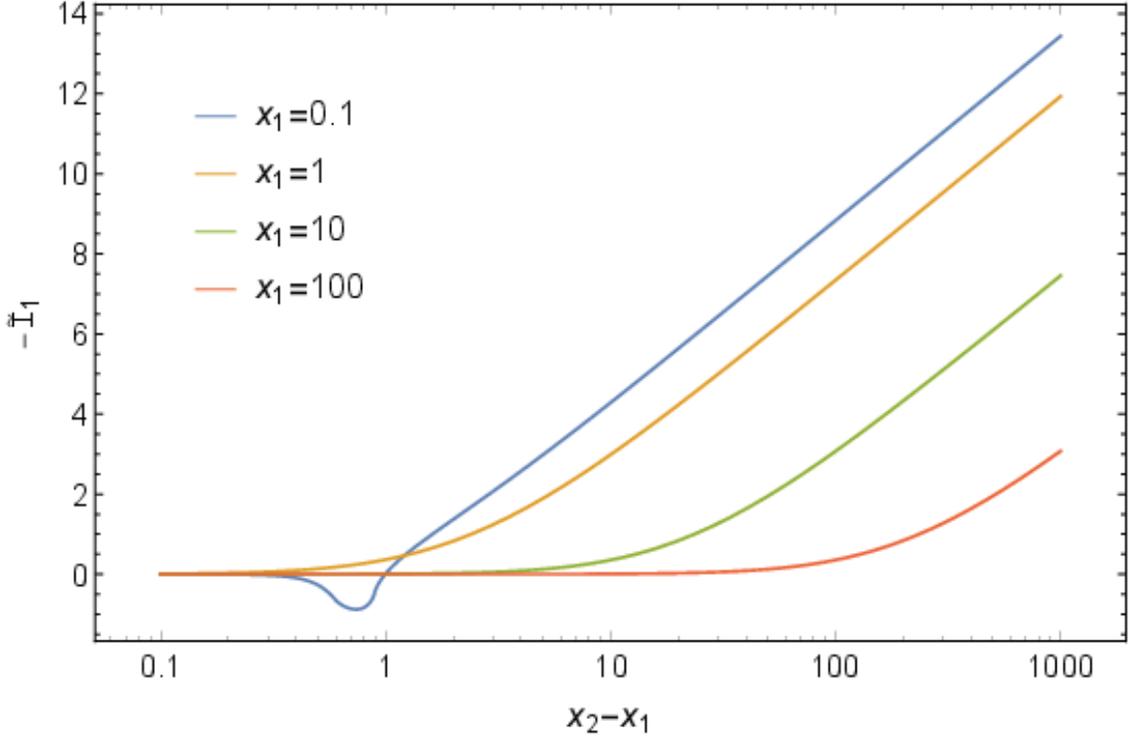}
  \caption{The one-loop integral $\tilde{I}_{1}^{}(x_1^{2},x_2^{2})\equiv I_{1R}^{}(x_1^{2},x_2^{2})$ or $\tilde{I}_{1}^{}(x_1^{2},x_2^{2})\equiv I_{1I}^{}(x_1^{2},x_2^{2})$ as a function of the dif\mbox{}ference $x_2^{}-x_1^{}$ for a given $x_1^{}$. The lines from top to bottom correspond to $x_1^{}=0.1$, $x_1^{}=1$, $x_1^{}=10$ and $x_1^{}=100$, respectively. }
  \label{integral1}
\end{figure*}

Furthermore, we can realize a $0\nu\beta\beta$ process at one-loop level. The relevant diagram is shown in Fig. \ref{nuless}b. The ef\mbox{}fective operator should be
\begin{widetext}
\begin{eqnarray}
\label{eff1loop}
\mathcal{O}^{\textrm{1-loop}}_{}&=& \frac{1}{\Lambda_1^5}\bar{u}_L^{}\gamma^\mu_{}d_L^{}
 \bar{u}_L^{}\gamma_\mu^{}d_L^{}\bar{e}_R^{}e_R^c~~\textrm{with}\nonumber\\
 [2mm]
\frac{1}{\Lambda_1^5}&=&\frac{1}{2^7_{}\pi^2}\frac{g^4_{}}{m_W^4}\frac{\omega}{m^2_{\xi^{\pm\pm}_{}}}f_{ee}^{}I_{1}^{}\sin^2_{}2\theta =64\,\pi^2_{}\,G_F^2 \frac{v^2_{}}{m_{\xi^{\pm\pm}_{}}^2}\frac{m_{ee}^{}}{m_e^2}\frac{I_1^{}}{I_2^{}}\frac{1}{c}\nonumber\\
[2mm]&=&9\times 10^4_{}\times \left(\frac{q}{100\,\textrm{MeV}}\right)^2_{} \left(\frac{1\,\textrm{TeV}}{m_{\xi^{\pm\pm}_{}}^{}}\right)^2_{}\left(\frac{I_1^{}}{1}\right)\left(\frac{1}{I_2^{}}\right)\left(\frac{1}{c}\right)
\frac{1}{\Lambda_2^5}\,.
\end{eqnarray}
\end{widetext}
Here $I_{1}^{}$ is given by the one-loop integral,
\begin{widetext}
\begin{eqnarray}
I_{1}^{}&=&I_{1R}^{}\left(\frac{m_{\hat{\chi}^\pm_{}}^2}{m_{\chi^{0}_{R}}^2},\frac{m_{\hat{\delta}^\pm_{}}^2}{m_{\chi^{0}_{R}}^2}\right)
-I_{1I}^{}\left(\frac{m_{\hat{\chi}^\pm_{}}^2}{m_{\chi^{0}_{I}}^2},\frac{m_{\hat{\delta}^\pm_{}}^2}{m_{\chi^{0}_{I}}^2}\right)\,,\nonumber\\
[2mm]
I_{1R}^{}\left(\frac{m_{\hat{\chi}^\pm_{}}^2}{m_{\chi^{0}_{R}}^2},\frac{m_{\hat{\delta}^\pm_{}}^2}{m_{\chi^{0}_{R}}^2}\right)&=&F_1^{}
\left(\frac{m_{\hat{\chi}^\pm_{}}^2}{m_{\chi^{0}_{R}}^2},\frac{m_{\hat{\chi}^\pm_{}}^2}{m_{\chi^{0}_{R}}^2}\right)
+F_1^{}
\left(\frac{m_{\hat{\delta}^\pm_{}}^2}{m_{\chi^{0}_{R}}^2},\frac{m_{\hat{\delta}^\pm_{}}^2}{m_{\chi^{0}_{R}}^2}\right)-2F_1^{}
\left(\frac{m_{\hat{\chi}^\pm_{}}^2}{m_{\chi^{0}_{R}}^2},\frac{m_{\hat{\delta}^\pm_{}}^2}{m_{\chi^{0}_{R}}^2}\right)\,,\nonumber\\
[2mm]
I_{1I}^{}\left(\frac{m_{\hat{\chi}^\pm_{}}^2}{m_{\chi^{0}_{I}}^2},\frac{m_{\hat{\delta}^\pm_{}}^2}{m_{\chi^{0}_{I}}^2}\right)&=&F_1^{}
\left(\frac{m_{\hat{\chi}^\pm_{}}^2}{m_{\chi^{0}_{I}}^2},\frac{m_{\hat{\chi}^\pm_{}}^2}{m_{\chi^{0}_{I}}^2}\right)+F_1^{}
\left(\frac{m_{\hat{\delta}^\pm_{}}^2}{m_{\chi^{0}_{I}}^2},\frac{m_{\hat{\delta}^\pm_{}}^2}{m_{\chi^{0}_{I}}^2}\right)-2F_1^{}
\left(\frac{m_{\hat{\chi}^\pm_{}}^2}{m_{\chi^{0}_{I}}^2},\frac{m_{\hat{\delta}^\pm_{}}^2}{m_{\chi^{0}_{I}}^2}\right)\,,
\end{eqnarray}
\end{widetext}
with the function $F_1^{}(a,b)$ being a double integral,
\begin{eqnarray}
F_1^{}(a,b)=\int_{0}^{1}dx\int_{0}^{1-x}dy \ln(1-x-y+ax+by)\,.
\end{eqnarray}
Clearly, $I_1^{}\equiv 0$ for $m_{\chi^0_R}^{}=m_{\chi^0_I}^{}$ or $m_{\hat{\xi}^{\pm}_{}}^{}=m_{\hat{\delta}^{\pm}_{}}^{}$. In Fig. \ref{integral1}, we show the numerical results of $\tilde{I}_{1}^{}(x_1^{2},x_2^{2})\equiv I_{1R}^{}(x_1^{2},x_2^{2})$ or $\tilde{I}_{1}^{}(x_1^{2},x_2^{2})\equiv I_{1I}^{}(x_1^{2},x_2^{2})$ as a function of the dif\mbox{}ference $ x_2^{}-x_1^{}$ for a given $x_1^{}$. The lines from top to bottom correspond to $x_1^{}=0.1$, $x_1^{}=1$, $x_1^{}=10$ and $x_1^{}=100$, respectively. The one-loop integral $I_1^{}=I_{1R}^{}-I_{1I}^{}$ can be of the order of $\mathcal{O}(1)$ for a proper parameter choice.

From Eqs. (\ref{eff2loop}), (\ref{efftree}) and (\ref{eff1loop}), we can conclude
\begin{eqnarray}
\frac{1}{\Lambda_{1}^5} >> \frac{1}{\Lambda_{01}^5} \,,~\frac{1}{\Lambda_{02}^5} \,,~\frac{1}{\Lambda_{03}^5} \,,~\frac{1}{\Lambda_{2}^5}\,,
\end{eqnarray}
for a reasonable parameter choice. The $0\nu\beta\beta$ thus should be dominated by the one-loop contribution. The lifetime is determined by \cite{hkk1996}
\begin{eqnarray}
\frac{1}{T_{1/2}^{0\nu}}&=&G^{}_{0\nu}|\eta_N^{} M_N^{}|^2_{}~~\textrm{with}~~
\eta_N^{}=\frac{16\pi^2_{}}{c}\frac{I_1^{}}{I_2^{}}\frac{m_{ee}^{} m_p^{} v^2_{}}{m_{\xi^{\pm\pm}_{}}^2 m_e^2}\,,~~
M_N^{}=\frac{m_p^{}}{m_e^{}}\left[\left(\frac{f_V^{}}{f_A^{}}\right)^2_{}M_{F,N}^{}-M_{GT,N}^{}\right]\,,
\end{eqnarray}
where $G^{}_{0\nu}$ is the phase space factor, $M_{F,N}^{}$ and $M_{GT,N}^{}$ are the nuclear matrix elements, while $f_{V}^{}\approx 1$ and $f_{A}^{}\approx 1.26$ normalize the hadronic current. For the $_{}^{136}\textrm{Xe}$ isotope with $G^{}_{0\nu}=3.56\times 10^{-14}_{}\,\textrm{yr}^{-1}_{}$, $M_{GT,N}^{}=0.058$ and $M_{F,N}^{}=-0.0203$, as well as $_{}^{76}\textrm{Ge}$ isotope with $G^{}_{0\nu}=5.77\times 10^{-15}_{}\,\textrm{yr}^{-1}_{}$, $M_{GT,N}^{}=0.113$ and $M_{F,N}^{}=-0.0407$ \cite{hkk1996}, we find
\begin{eqnarray}
\label{prediction}
T_{1/2}^{0\nu}(_{}^{136}\textrm{Xe})&=&1.25\times 10^{26}_{}\,\textrm{yr}\times \left(\frac{10^{-7}_{}\,\textrm{eV}}{|m_{ee}^{}|}\right)^2_{}\left(\frac{m_{\xi^{\pm\pm}}}{1\,\textrm{TeV}}\right)^4_{}\left(\frac{c}{1}\right)^2_{}\left(\frac{I_2^{}}{1}\right)^2
\left(\frac{1}{I_1^{}}\right)^2_{}\,,\nonumber\\
[2mm]
T_{1/2}^{0\nu}(_{}^{76}\textrm{Ge})&=&2.0\times 10^{26}_{}\,\textrm{yr}\times \left(\frac{10^{-7}_{}\,\textrm{eV}}{|m_{ee}^{}|}\right)^2_{}\left(\frac{m_{\xi^{\pm\pm}}}{1\,\textrm{TeV}}\right)^4_{}\left(\frac{c}{1}\right)^2_{}\left(\frac{I_2^{}}{1}\right)^2
\left(\frac{1}{I_1^{}}\right)^2_{}\,.
\end{eqnarray}
Currently the experimental limits are $T_{1/2}^{0\nu}(_{}^{136}\textrm{Xe})>1.07\times 10^{26}_{}\,\textrm{yr}$ from the KamLAND-Zen collaboration \cite{shirai2016} and $T_{1/2}^{0\nu}(_{}^{76}\textrm{Ge})>5.2\times 10^{25}_{}\,\textrm{yr}$ from the GERDA collaboration \cite{agostini2016}. The $0\nu\beta\beta$ half-life sensitivity is expected to improve in the future, such as $T_{1/2}^{0\nu}(_{}^{136}\textrm{Xe})>8\times 10^{26}_{}\,\textrm{yr}$ \cite{auger2012} and $T_{1/2}^{0\nu}(_{}^{76}\textrm{Ge})>6\times 10^{27}_{}\,\textrm{yr}$ \cite{ackermann2013,abgrall2013}.

\section{Other constraints and implications}

The Yukawa couplings of the doubly charged scalar $\xi^{\pm\pm}_{}$ to the right-handed leptons $(e_{R}^{\pm},\,\mu_R^{\pm},\,\tau_R^{\pm})$ can result in other experimental implications \cite{nops2008,hnrs2014} such as $e^{+}_{}e^{-}_{}\rightarrow e^{+}_{}e^{-}_{}$, $e^{+}_{}e^{-}_{}\rightarrow \mu^{+}_{}\mu^{-}_{}$, $\mu^{-}_{}e^{+}_{} \rightarrow \mu^{+}_{}e^{-}_{}$, $\mu\rightarrow 3e$, $\mu\rightarrow e\gamma$, $(g-2)_{\mu}^{}$ and so on. By integrating out the doubly charged scalar and then using the Fierz transformation, we can easily give the ef\mbox{}fective Lagrangian for the electron-positron reactions $e^{+}_{}e^{-}_{}\rightarrow e^{+}_{}e^{-}_{}$ and $e^{+}_{}e^{-}_{}\rightarrow \mu^{+}_{}\mu^{-}_{}$, i.e.
\begin{eqnarray}
\mathcal{L}&\supset&\frac{|f_{ee}^{}|^2_{}}{8m_{\xi^{\pm\pm}_{}}^2}(\bar{e}_R^{}\gamma^\mu_{}e_R^{})( \bar{e}_R^{}\gamma_\mu^{} e_R^{})
+\frac{|f_{e\mu}^{}|^2_{}}{2m_{\xi^{\pm\pm}_{}}^2}(\bar{e}_R^{}\gamma^\mu_{}e_R^{})( \bar{\mu}_R^{}\gamma_\mu^{} \mu_R^{})\nonumber\\
[2mm]&=&\frac{2^{13}_{}\pi^{8}_{}}{c^2_{}I_2^2 \sin^4_{}2\theta}\frac{v^4_{}|m_{ee}^{}|^2_{}}{m_e^4 m_{\xi^{\pm\pm}_{}}^2\omega^2_{} }(\bar{e}_R^{}\gamma^\mu_{}e_R^{})( \bar{e}_R^{}\gamma_\mu^{} e_R^{})
+ \frac{2^{15}_{}\pi^{8}_{}}{c^2_{}I_2^2 \sin^4_{}2\theta}\frac{v^4_{}|m_{e\mu}^{}|^2_{}}{m_e^2 m_\mu^2 m_{\xi^{\pm\pm}_{}}^2\omega^2_{} }(\bar{e}_R^{}\gamma^\mu_{}e_R^{})( \bar{\mu}_R^{}\gamma_\mu^{} \mu_R^{})\,.
\end{eqnarray}
Similarly, we can give the ef\mbox{}fective Lagrangian for the muonium-antimuonium conversion $\mu^{-}_{}e^{+}_{} \rightarrow \mu^{+}_{}e^{-}_{}$ as below,
\begin{eqnarray}
\mathcal{L}\supset\frac{f_{ee}^{}f_{\mu\mu}^{\ast}}{8m_{\xi^{\pm\pm}_{}}^2}(\bar{e}_R^{}\gamma^\mu_{}\mu_R^{})( \bar{e}_R^{}\gamma_\mu^{} \mu_R^{})+\textrm{H.c.}=\frac{2^{13}_{}\pi^{8}_{}}{c^2_{}I_2^2 \sin^4_{}2\theta}\frac{v^4_{} m_{ee}^{} m_{\mu\mu}^\ast}{m_e^2 m_\mu^2 m_{\xi^{\pm\pm}_{}}^2\omega^2_{} }(\bar{e}_R^{}\gamma^\mu_{}\mu_R^{})( \bar{e}_R^{}\gamma_\mu^{} \mu_R^{})+\textrm{H.c.}\,.
 \end{eqnarray}
For the rare three-body decay $\mu\rightarrow 3e$, it can be described by 
the ef\mbox{}fective Lagrangian, 
\begin{eqnarray}
\mathcal{L}\supset\frac{f_{ee}^{}f_{e\mu}^{\ast}}{2m_{\xi^{\pm\pm}_{}}^2}(\bar{e}_R^{}\gamma^\mu_{}e_R^{})( \bar{e}_R^{}\gamma_\mu^{} \mu_R^{})+\textrm{H.c.}=\frac{2^{15}_{}\pi^{8}_{}}{c^2_{}I_2^2 \sin^4_{}2\theta}\frac{v^4_{} m_{ee}^{}m_{e\mu}^{\ast}}{m_e^3 m_\mu^{} m_{\xi^{\pm\pm}_{}}^2\omega^2_{} }(\bar{e}_R^{}\gamma^\mu_{}e_R^{})( \bar{e}_R^{}\gamma_\mu^{} \mu_R^{})+\textrm{H.c.}\,,
\end{eqnarray}
and then its decay width can be computed by 
\begin{eqnarray}
\Gamma_{\mu\rightarrow 3 e}=\frac{|f_{e\mu}^{} f_{e e}^{}|^2_{}}{3\times 2^{12}_{}\pi^3_{}}\frac{m_\mu^5}{m_{\xi^{\pm\pm}_{}}^4}=\frac{2^{20}_{}\pi^{13}_{}}{3\,c^4_{}I_2^4 \sin^8_{}2\theta}\frac{v^8_{}m_\mu^3|m_{ee}^{}|^2_{}|m_{e\mu}^{}|^2_{}}{m_e^6 m_{\xi^{\pm\pm}_{}}^4  \omega^4_{}}\,.
\end{eqnarray}
By taking into account the SM result of the muon total decay width, 
\begin{eqnarray}
\Gamma_{\mu}^{}\simeq \Gamma_{\mu\rightarrow e\bar{\nu}_e^{} \nu_\mu^{}}^{\textrm{SM}}=\frac{G_F^2 m_\mu^5}{192\,\pi^3_{}}\,,
\end{eqnarray}
we can read the branching ratio,
\begin{eqnarray}
\label{mu3e}
Br(\mu\rightarrow 3 e)=\frac{\Gamma_{\mu\rightarrow 3 e}}{\Gamma_{\mu}^{}}=\frac{2^{26}_{}\pi^{16}_{}}{c^4_{}I_2^4 \sin^8_{}2\theta}\frac{v^8_{}|m_{ee}^{}|^2_{}|m_{e\mu}^{}|^2_{}}{G_F^2 m_e^6 m_\mu^{2} m_{\xi^{\pm\pm}_{}}^4 \omega^4_{} }\,.
\end{eqnarray}
Clearly, we can get the similar formula for the other rare three-body decays $\tau^{-}_{}\rightarrow 3\mu^{}$, $\mu^{+}_{}\mu^{-}_{}e^{-}_{}$, $e^{+}_{}\mu^{-}_{}\mu^{-}_{}$, $\mu^{-}_{} e^{+}_{}e^{-}_{}$, $\mu^{+}_{}e^{-}_{}e^{-}_{}$, $3e$ by replacing the related parameters $m_{\alpha\beta}^{}$ in Eq. (\ref{mu3e}). We also consider the lepton flavor changing decay $\mu \rightarrow e \gamma$. The decay width should be 
\begin{eqnarray}
\Gamma_{\mu\rightarrow e\gamma}&=&\frac{\alpha m_\mu^5}{9\times 2^{12}_{}\,\pi^3_{}}\frac{|f_{\mu e}^{} f_{e e}^{}|^2_{}+|f_{\mu e}^{} f_{\mu\mu}^{}|^2_{}+4|f_{\mu \tau}^{} f_{e \tau}^{}|^2_{}}{m_{\xi^{\pm\pm}_{}}^4}\nonumber\\
[2mm]&=&\frac{2^{20}_{}\pi^{13}_{}\alpha}{9\,c^4_{}I_2^4 \sin^8_{}2\theta}\frac{v^8_{}}{ m_{\xi^{\pm\pm}_{}}^4  \omega^4_{}}\left(\frac{m_\mu^3|m_{ee}^{}|^2_{}|m_{e\mu}^{}|^2_{}}{m_e^6}+
\frac{|m_{e\mu}^{}|^2_{}|m_{\mu\mu}^{}|^2_{}}{m_e^2 m_\mu^{}}+4\frac{m_\mu^3|m_{e\tau}^{}|^2_{}|m_{\mu\tau}^{}|^2_{}}{m_e^2 m_\tau^4}\right)\,,
\end{eqnarray}
and then the branching ratio,
\begin{eqnarray}
Br(\mu\rightarrow e\gamma)=\frac{\Gamma_{\mu\rightarrow e \gamma}}{\Gamma_{\mu}^{}}=\frac{2^{26}_{}\pi^{16}_{}\alpha}{3\,c^4_{}I_2^4 \sin^8_{}2\theta}\frac{v^8_{}}{G_F^2 m_{\xi^{\pm\pm}_{}}^4 \omega^4_{} }\left(\frac{|m_{ee}^{}|^2_{}|m_{e\mu}^{}|^2_{}}{m_e^6 m_\mu^2}+
\frac{|m_{e\mu}^{}|^2_{}|m_{\mu\mu}^{}|^2_{}}{m_e^2 m_\mu^{6}}+4\frac{|m_{e\tau}^{}|^2_{}|m_{\mu\tau}^{}|^2_{}}{m_e^2 m_\mu^2 m_\tau^4}\right)\,.
\end{eqnarray}
For simplicity, we do not show the similar formula for the other lepton flavor changing decays $\tau \rightarrow \mu \gamma$, $e\gamma$. We then calculate the muon anomalous magnetic momnet $(g-2)_{\mu}^{}$, i.e.
\begin{eqnarray}
\Delta a_\mu^{}=-\frac{m_\mu^2}{96\pi^2_{}}\frac{4|f_{\mu e}^{}|^2_{}+|f_{\mu\mu}^{}|^2_{}+4|f_{\mu\tau}^{}|^2_{}}{m_{\xi^{\pm\pm}_{}}^2}
=-\frac{2^{11}_{}\pi^{6}_{}}{3\,c^2_{}I_2^2 \sin^4_{}2\theta}\frac{v^4_{}}{ m_{\xi^{\pm\pm}_{}}^2 \omega^2_{}  }\left(4\frac{|m_{e\mu}^{}|^2_{}}{m_e^2}+
\frac{|m_{\mu\mu}^{}|^2_{}}{m_\mu^{2}}+4\frac{|m_{\mu\tau}^{}|^2_{}}{m_\tau^2}\right)\,.
\end{eqnarray}
We have checked that our model can escape from the above experimental constraints \cite{patrignani2016} for the parameter choice $c=\mathcal{O}(1)$, $I_2^{}=\mathcal{O}(1)$, $\sin 2 \theta\leq 1$, $m_{\xi^{\pm\pm}_{}}^{}=\mathcal{O}(\textrm{TeV})$, $\omega =\mathcal{O}( \textrm{TeV})$, $m_{e\mu,e\tau,\mu\mu,\mu\tau,\tau\tau}^{}=\mathcal{O}(0.1\,\textrm{eV})$ and $m_{ee}^{}\lesssim\mathcal{O}(10^{-4}_{}\,\textrm{eV})$.

\begin{figure*}
\vspace{8.5cm} \epsfig{file=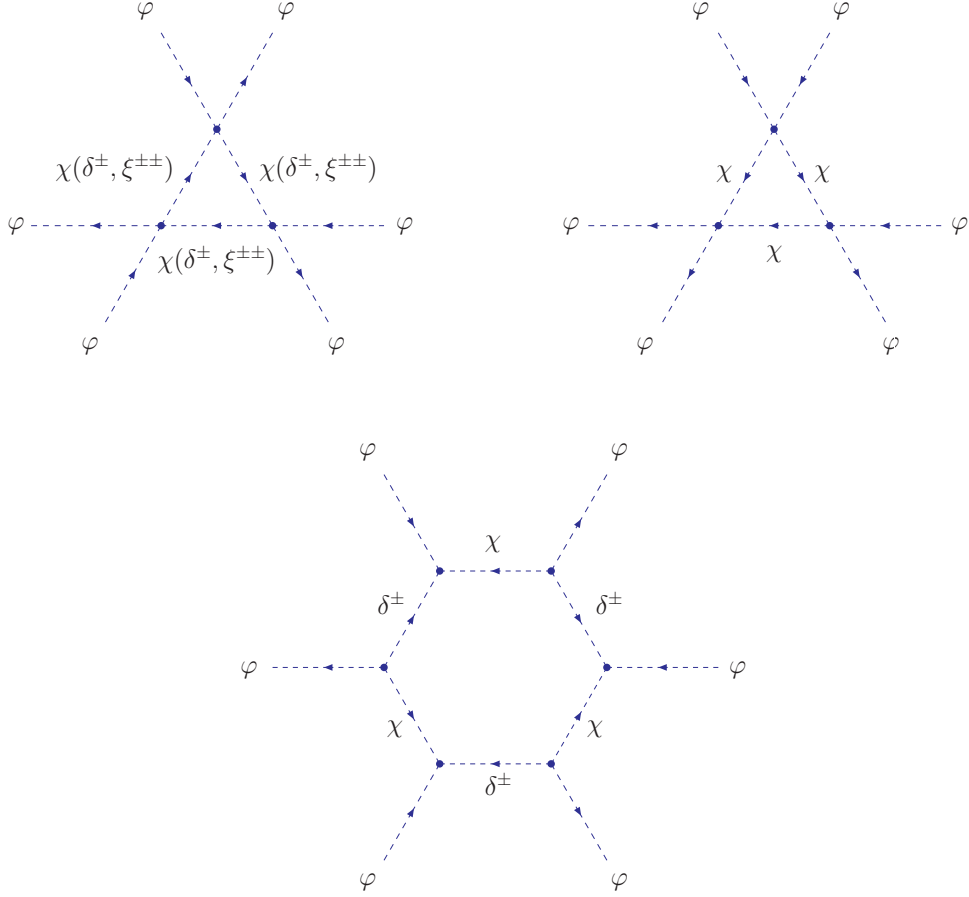, bbllx=5cm,
bblly=6.0cm, bburx=15cm, bbury=16cm, width=7cm, height=7cm,
angle=0, clip=0} \vspace{-3cm} \caption{\label{higgs}
One-loop diagrams for giving a dimension-6 term of the SM Higgs scalar.} 
\end{figure*}

We further consider the couplings of the non-SM scalars including the doublet $\chi$ as well as the singlets $\delta^{\pm}_{}$ and $\xi^{\pm\pm}_{}$ to the SM Higgs doublet $\varphi$. As shown in Fig. \ref{higgs}, these non-SM scalars can mediate some one-loop diagrams to give a dimension-6 operator \cite{zhang1993} of the SM Higgs doublet $\varphi$. The Higgs potential thus should be 
\begin{eqnarray}
\label{dim6}
 V &\supset&\mu_\varphi^2\varphi^\dagger_{}\varphi+\kappa_1^{}(\varphi^\dagger_{}\varphi)^2_{}+ \frac{1}{\Lambda^2_{6}}(\varphi^\dagger_{}\varphi)^3_{}~~\textrm{with}\nonumber\\
[2mm]
\frac{1}{\Lambda^2_{6}}&=&\frac{1}{16\pi^2_{}}\left\{\frac{\kappa_3^3+(\kappa_3^{}+\kappa_4^{})^3_{}+(\kappa_3^{}+\kappa_4^{})|\kappa_5^{}|^2_{}}{M_\chi^2}+\frac{\kappa_{\varphi\delta}^{3}}{M_\delta^2}+\frac{\kappa_{\varphi\xi}^{3}}{M_\xi^2}\right.\nonumber\\
[2mm]&&\left.+
\frac{\sigma^6_{}}{(M_\chi^2-M_\delta^2)^4_{}}\left[\frac{M_\chi^4+10M_\chi^2 M_\delta^2 + M_\delta^4}{3M_\chi^2 M_\delta^2 }+2\frac{M_\chi^2+M_\delta^2}{M_\chi^2-M_\delta^2}\ln\left(\frac{M_\delta^2}{M_\chi^2}\right)\right]
\right\}\,.
\end{eqnarray}
By minimizing this potential, the quadratic and trilinear terms of the Higgs boson $h$ can be extracted, 
\begin{eqnarray}
\mathcal{L}&\supset& -\frac{1}{2}m_h^2 h^2_{}-\kappa_{\textrm{ef\mbox{}f}}^{} v h^3_{}~~\textrm{with}~~m_h^2=2\kappa_1^{} v^2_{} +3 \frac{v^4_{}}{\Lambda_6^2}\,,~~\kappa_{\textrm{ef\mbox{}f}}^{}=\kappa_1^{}+\frac{5v^2_{}}{2\Lambda_6^2}
=\frac{m_h^2}{2v^2_{}}+\frac{v^2_{}}{\Lambda_6^2}\,.
\end{eqnarray}
The trilinear coupling of the Higgs boson yields a deviation from its SM value,
\begin{eqnarray}
R_{\lambda}^{}=\frac{\lambda_{\textrm{ef\mbox{}f}}^{}-\lambda_{\textrm{SM}}^{}}{\lambda_{\textrm{SM}}^{}}
=\frac{2v^4_{}}{m_h^2 \Lambda_6^2 }=0.12\left(\frac{2\,\textrm{TeV}}{\Lambda_6^{}}\right)^2_{}~~\textrm{with}~~\lambda_{\textrm{SM}}^{}=\frac{m_h^2}{2v^2_{}}\,.
\end{eqnarray}
The dimension-6 operator (\ref{dim6}) may have an interesting ef\mbox{}fect on the electroweak phase transition and hence the electroweak baryogenesis \cite{zhang1993} and may be tested at the running and future colliders\cite{mccullough2014,em2013,hgyyz2015,hjlw2016,dhot2016,zlly2016,hlw2016,klwy2016}. We now check the Higgs to diphoton decay \cite{clw2012,englert2014,cordero-cid2014,am2012,arhrib2014},
\begin{eqnarray}
R_{\gamma\gamma}^{}&\equiv& \frac{\Gamma\left(h\rightarrow \gamma\gamma\right)}{\Gamma_{\textrm{SM}}\left(h\rightarrow \gamma\gamma\right)} \nonumber\\
[2mm]
&=&\left|1+\frac{\kappa_{\hat{\chi}^{\pm}_{}}^{}v_{}^2}{2 m_{\hat{\chi}^\pm_{}}^2}\frac{A_0^{}\left(\tau_{\hat{\chi}^\pm_{}}^{}\right)}{A_1^{}\left(\tau_W^{} \right)+\frac{4}{3} A_{\frac{1}{2}}^{}\left(\tau_t^{}\right)}
+\frac{\kappa_{\hat{\delta}^{\pm}_{}}^{}v_{}^2}{2 m_{\hat{\delta}^\pm_{}}^2}\frac{A_0^{}\left(\tau_{\hat{\delta}^\pm_{}}^{}\right)}{A_1^{}\left(\tau_W^{} \right)+\frac{4}{3} A_{\frac{1}{2}}^{}\left(\tau_t^{}\right)}
+\frac{2\kappa_{\varphi\xi}^{}v_{}^2}{m_{\xi^{\pm\pm}_{}}^2}\frac{A_0^{}\left(\tau_{\xi^{\pm\pm}_{}}^{}\right)}{A_1^{}\left(\tau_W^{} \right)+\frac{4}{3} A_{\frac{1}{2}}^{}\left(\tau_t^{}\right)}
\right|^2\nonumber\\
[2mm]&&\textrm{with}~~\kappa_{\hat{\chi}^{\pm}_{}}^{}=\kappa_3^{}\cos^2_{}\theta+\kappa_{\varphi\delta}^{}\sin^2_{}\theta-\frac{\sigma\sin 2\theta }{\sqrt{2} v}\,,~~\kappa_{\hat{\delta}^{\pm}_{}}^{}=\kappa_3^{}\sin^2_{}\theta+\kappa_{\varphi\delta}^{}\cos^2_{}\theta+\frac{\sigma\sin 2\theta }{\sqrt{2} v}\,,~~\tau_X^{}=4\frac{m_X^2}{m_h^2}\,,\nonumber\\
[2mm]&&~~~~~~~\,A_{0}^{}(x)=-x^2_{}\left(\frac{1}{x}-\arcsin^2_{}\frac{1}{\sqrt{x}}\right)\,,~~A_{1}^{}(x)=-x^2_{}\left[\frac{2}{x^2_{}}+\frac{3}{x}+3\left(\frac{2}{x}-1\right)\arcsin^2_{}\frac{1}{\sqrt{x}}\right]\,,\nonumber\\
[2mm]&&~~~~~~~\,A_{\frac{1}{2}}^{}(x)=2x^2_{}\left[\frac{1}{x}+\left(\frac{1}{x}-1\right)\arcsin^2_{}\frac{1}{\sqrt{x}}\right]\,.
\end{eqnarray}
We find for $\kappa_{\hat{\chi}^{\pm}_{}}^{}=\mathcal{O}(1)$, $\kappa_{\hat{\delta}^{\pm}_{}}^{}=\mathcal{O}(1)$, $\kappa_{\varphi\xi}^{}=\mathcal{O}(1)$, $m_{\hat{\chi}^{\pm}_{}}^{}=\mathcal{O}(\textrm{TeV})$, $m_{\hat{\delta}^{\pm}_{}}^{}=\mathcal{O}(\textrm{TeV})$ and $m_{\xi^{\pm\pm}_{}}^{}=\mathcal{O}(\textrm{TeV})$, the value of $R_{\gamma\gamma}^{}$ can be much below the experimental limit \cite{patrignani2016}.

\section{Summary}

In this paper we have demonstrated the Majorana neutrino mass generation in some extended two Higgs doublet models, where the Yukawa couplings for the charged fermion mass generation only involve one Higgs scalar. In our scenario, a singly charged scalar without any Yukawa couplings has a cubic term with the two Higgs scalars. Another doubly charged scalar has the Yukawa couplings with the right-handed leptons, besides its trilinear coupling with the singly charged scalar. After the electroweak symmetry breaking, we can obtain the desired neutrino masses at two-loop level. The symmetric Yukawa couplings of the doubly charged scalar to the right-handed leptons can fully determine the structure of the neutrino mass matrix. Meanwhile, a one-loop $0\nu\beta\beta$ process can arrive at an observable level even if the electron neutrino mass is extremely tiny. We have also checked the other experimental constraints and implications including some rare processes and Higgs phenomenology.

\textbf{Acknowledgement}: The authors were supported by the Recruitment Program for Young Professionals under Grant No. 15Z127060004, the Shanghai Jiao Tong University under Grant No. WF220407201 and the Shanghai Laboratory for Particle Physics and Cosmology under Grant No. 11DZ2260700. This work was also supported by the Key Laboratory for Particle Physics, Astrophysics and Cosmology, Ministry of Education.

\end{document}